\documentclass[journal]{IEEEtran}

\ifCLASSINFOpdf
\else
\fi

\usepackage{cite}
\usepackage[pdftex]{graphicx}
\setkeys{Gin}{clip=true,draft=false}
\DeclareGraphicsExtensions{.pdf}
\usepackage[cmex10]{amsmath}
\usepackage{amsthm}
\usepackage{amsfonts}
\usepackage[cspex,bbgreekl]{mathbbol}
\usepackage[hang]{subfigure}
\usepackage{color}
\usepackage[cmex10]{amsmath}
\usepackage{amssymb, amsthm}
\usepackage{mathtools}
\usepackage{textcomp}

\makeatother

\begin{document}
\title{Stress-testing memcomputing on hard combinatorial optimization problems}

\author{Forrest Sheldon, Pietro Cicotti, Fabio L. Traversa, Massimiliano Di Ventra\thanks{F. Sheldon and M. Di Ventra are with the Department of Physics, University of California-San Diego, 9500  Gilman Drive, La Jolla, California 92093-0319, USA. P. Cicotti is with the San Diego Supercomputer Center, La Jolla, 92093 CA. F.L. Traversa is with MemComputing Inc., San Diego, CA, 92130 CA. e-mail: fsheldon@ucsd.edu, ftraversa@memcpu.com, pcicotti@sdsc.edu, diventra@physics.ucsd.edu}}

\maketitle

\begin{abstract}
Memcomputing is a novel paradigm of computation that utilizes dynamical elements with memory to both store and process information on the same physical location. Its building blocks can be fabricated in hardware with standard electronic circuits, thus offering a path to its practical realization. In addition, since memcomputing is based on non-quantum elements, the equations of motion describing these machines can be \emph{simulated} efficiently on standard computers. In fact, it was recently realized that memcomputing, and in particular its {\it digital} (hence scalable) version, when simulated on a classical machine provides a significant speed-up over state-of-the-art algorithms on a variety of non-convex problems. Here, we stress-test the capabilities of this approach on finding approximate solutions to hard combinatorial optimization problems.   These fall into a class which is known to require exponentially growing resources in the worst cases, even to generate approximations.  We recently showed that in a region where state of the art algorithms demonstrate this exponential growth, simulations of digital memcomputing machines performed using the Falcon$^\copyright$ simulator of MemComputing, Inc. only require time and memory resources that scale {\it linearly}.   These results are extended in a stress-test up to $64\times10^6$ variables (corresponding to about 1 billion literals), namely the largest case that we could fit on a single node with 128 GB of DRAM. Since memcomputing can be applied to a wide variety of optimization problems, this stress test shows the considerable advantage of non-combinatorial, physics-inspired approaches over standard combinatorial ones.
\end{abstract}

%
%

\maketitle

\section{Introduction}
The increasing demand for computational power and efficiency is driving the scientific community and industry to explore new and unconventional ways to compute. In this respect, new ideas and radically different paradigms may be the key to solve or mitigate the computational bottlenecks that affect 
present computing technology.   

A new computing paradigm has been recently proposed called {\it memcomputing} (which stands for computing {\it in} and {\it with} memory)~\cite{13_memcomputing,UMM}, with the potential to increase the computational efficiency in the solution of hard combinatorial/optimization problems. The formal description of memcomputing rests on the concept of Universal Memcomputing Machines (UMMs)~\cite{UMM}. UMMs are a collection of interconnected memprocessors (processors with memory) able to process and store information on the same physical location. UMMs are a class of non-Turing machines with specific properties as described in \cite{UMM,ONUMM}. In particular, they support intrinsic parallelism, i.e., interconnected (mem-)processors act collectively on data using their collective state to perform computation\cite{UMM,DMM2,traversaNP}. Moreover, memprocessors are able to exploit the information available through the {\it topology} of their connections. Indeed, specific network topologies may be designed to embed problems in a one-to-one mapping. This last property has been named {\it information overhead}~\cite{UMM}. 

We have recently shown that the {\it digital} (hence scalable) subclass of UMMs we call {\it digital memcomputing machines} (DMMs) can deliver substantial 
benefits in the solution of hard combinatorial/optimization problems compared to traditional algorithmic approaches. For instance, in Ref.~\cite{DMMperspective} we have shown that the simulation of DMMs on 
a classical computer solves the search version of the subset-sum problem in polynomial time for the worst cases. In Ref.~\cite{AcceleratingDL} we have  
shown how to accelerate the pre-training of restricted Boltzmann machines using simulations of DMMs, and demonstrated their advantage over quantum-based machines (implemented in hardware), such as D-wave machines~\cite{Adachi2015}, as well as state-of-the-art supervised learning~\cite{glorot2011deep,pmlr-v37-ioffe15}. Finally, in Ref.~\cite{exponential2017speedup}, we have employed simulations of DMMs and shown substantial advantages over traditional algorithms for a wide variety of optimization problems such as the Random 2 Max-SAT, the Max-Cut, the Forced Random Binary problem, and the Max-Clique~\cite{complexity_bible}. In some cases, the memcomputing approach obtains the solution to problems on which the winners of the 2016 Max-SAT competition failed~\cite{exponential2017speedup}.

We have also  performed scalability tests on finding approximate solutions to hard constructed instances of the Max-SAT problem~\cite{exponential2017speedup}.  Max-SAT possesses an \emph{inapproximability gap}~\cite{Hastad2001}, meaning that even calculating an approximate solution beyond a certain fraction of the optimum will require resources which grow exponentially with the input size in the worst case.   We observed this exponential growth on the tested instances with some of the best solvers from the 2016 Max-SAT competition.  Instead, our simulations done for these hard cases, and up to a 
certain problem size  
succeed in overcoming that gap in linear time employing memory (of the single processor used for the simulations) that also scales linearly with input size~\cite{exponential2017speedup}.

Our goal for the present paper is to understand how far we can push these simulations with the available resources at our disposal. We focus 
again on hard combinatorial optimization problems since these arise in nearly every industrial and scientific discipline. These problems involve the minimization or maximization of a function of many independent variables, often called the cost function, whose value represents the quality of a given solution~\cite{Optimization_book,Optimization_book_intro}. In industrial settings this may be the wiring cost of a computer chip layout, or the length of a delivery route~\cite{kirkpatricksimanneal}.  In scientific applications it may be searching for the ground state of a spin system or proteins~\cite{kobe2006}.   

We will show here that the simulations of DMMs on hard optimization problems can be pushed to tens of millions of variables, corresponding to about one billion literals, using a commercial and sequential MATLAB code (the Falcon$^\copyright$ simulator provided by MemComputing, Inc.) running on a single thread of an Intel Xeon E5-2680 v3 with 128 Gb DRAM shared on 24 threads. These results show once more the power of physics-inspired approaches to computation over traditional algorithmic methods. 

\section{The problem}
In this work, we focus on cost functions over a set of Boolean variables $x_1, \dots, x_N,$ $x_i \in \{0,1\}$, into which a large number of problems may be cast.

A common formulation of these problems is given by a set of Boolean constraints where the cost function counts the number of constraints satisfied by an assignment.  This is often expressed in conjunctive normal form (CNF) where each constraint (also called a \emph{clause}) is expressed as the disjunction (logical OR denoted by $\vee$) of a set of \emph{literals} (a variable $x_i$ or its negation $\overline{x_i}$) which are then conjoined (logical AND denoted by $\wedge$) together, e.g., in expressions of the type:
\begin{align*}
(x_1 \vee x_2)  \wedge (x_1 \vee \overline{x_2} \vee  x_3) \wedge (x_1 \vee \overline{x_2} \vee \overline{x_3} ) \wedge (\overline{x_1} \vee \overline{x_2} \vee x_3) \\
{}\wedge (\overline{x_1} \vee  \overline{x_3}) \wedge (\overline{x_2} \vee x_3)\wedge (\overline{x_1} \vee x_2 \vee x_3) .
\end{align*}
The CNF representation may be considered general in the sense that any Boolean formula may be expressed in this form~\cite{complexity_bible}. 

The problem of determining the assignment satisfying the maximum number of clauses is known as Max-SAT and is NP-hard, i.e., any problem in the class non-deterministic polynomial (NP) may be reduced to it in polynomial time~\cite{complexity_bible}. As a result, algorithms for obtaining the solution will, in the worst cases, require a time that scales exponentially with the size of the instance, thus creating severe bottlenecks in the solution of complex industrial and scientific problems.  The ubiquity and importance of this problem is exemplified by the yearly MAX-SAT competitions, where state-of-the-art solvers are tested on a variety of benchmarks to stimulate research and innovation~\cite{MAXSAT_competition}.

In applications where the optimal solution is required \emph{exact} algorithms must be used~\cite{gomes2008satisfiability}.  These {\it complete} solvers typically proceed by first obtaining bounds on the quality of the solution and then using these bounds to prune the search tree in a backtracking search.  This systematic approach guarantees that the resulting solution will be the optimum, but typically scales poorly and is impractical for large instances.

In these cases, \emph{heuristic} or \emph{incomplete} algorithms must be used~\cite{gomes2008satisfiability,kautz2009incomplete,heuristics_book}.  Rather than systematically searching the solution space, these solvers generate an initial assignment and then iteratively improve upon it, using a variety of strategies to boost the efficiency of their local search. After a specified number of steps, the algorithm returns its best assignment. As randomness is often used to drive the search, this procedure is referred to as \emph{stochastic local search}. While they can no longer guarantee optimatility, these solvers have proven very effective at approximating, and sometimes solving difficult Max-SAT instances.  For instance, in the 2016 Max-SAT competition~\cite{MAXSAT_competition}, incomplete solvers performed 2 orders of magnitude faster than complete solvers on random and crafted benchmarks.

We might hope that if we seek an \emph{approximation}, rather than a solution of a Max-SAT instance, we could avoid the exponential scaling of the run-time. Unfortunately, it turns out that even approximating the solution of many difficult optimization problems is NP-hard.  More precisely, for a maximization problem with optimum $O$ defined as the sum of the weights of all satisfied clauses, obtaining an assignment better than $f O$ for the fraction $f$ greater than some critical fraction, $f_c$, is an NP-hard problem~\cite{Feige1998,Hastad2001}.  For example, obtaining an assignment for Max-E3SAT (a version of Max-SAT in which every clause has 3 literals) better than $f_c = 7/8$ of the optimum is NP-hard, meaning that we cannot expect a polynomial algorithm to obtain the approximation for any instance of Max-E3SAT unless NP=P. Any known algorithm thus must show exponential scaling for a threshold past $f_c$ in the worst case. In~\cite{exponential2017speedup} we showed a region in which some of the best algorithms based on stochastic local search show an exponential growth with input size, but where the memcomputing approach based on deterministic dynamical systems requires only linearly growing time to achieve the same threshold.

\section{Digital Memcomputing Machines}

As mentioned in the Introduction, we present here a radically novel approach to these problems based on the simulation of DMMs on a standard classical computer~\cite{DMM2,UMM}.  DMMs can be implemented in practice as specially-designed dynamical systems whose equilibrium (fixed) points represent the approximations to the computational problem at hand, and which can be realized with standard electrical components and those with memory.

The mathematical definition of DMMs is the eight-tuple~\cite{DMM2}
\begin{equation}
	DMM=(\mathbb{Z}_2,\Delta,{\mathcal P},S,\Sigma,p_0,s_0,F)\,,\label{UMMdef}
\end{equation}
where (although not strictly necessary) we consider the range of $\mathbb{Z}_2=\{0,1\}$. Generalization to any finite number of states is trivial. $\Delta$ is a set of functions
\begin{equation}
	\delta_\alpha:\mathbb{Z}_2^{m_\alpha}\backslash F\times {\mathcal P}\rightarrow \mathbb{Z}_2^{m\rq_\alpha}\times {\mathcal P}^2\times S\,,\label{functUMM}
\end{equation}
where $m_\alpha<\infty$ is the number of memprocessors used as input of (read by) the function $\delta_\alpha$, and $m\rq_\alpha<\infty$ is the number of memprocessors used as output (written by) the function $\delta_\alpha$; ${\mathcal P}$ is the set of the arrays of pointers $p_\alpha$ that select the memprocessors called by $\delta_\alpha$ and $S$ is the set of indexes $\alpha$; $\Sigma$ is the set of the initial states written by the input device on the computational memory; $p_0\in {\mathcal P}$ is the initial array of pointers; $s_0$ is the initial index $\alpha$,  and $F\subseteq \mathbb{Z}_2^{m_f}$ for some $m_f\in\mathbb{N}$ is the set of final states.
\begin{figure*}[t]
	\includegraphics[width=1.8\columnwidth]{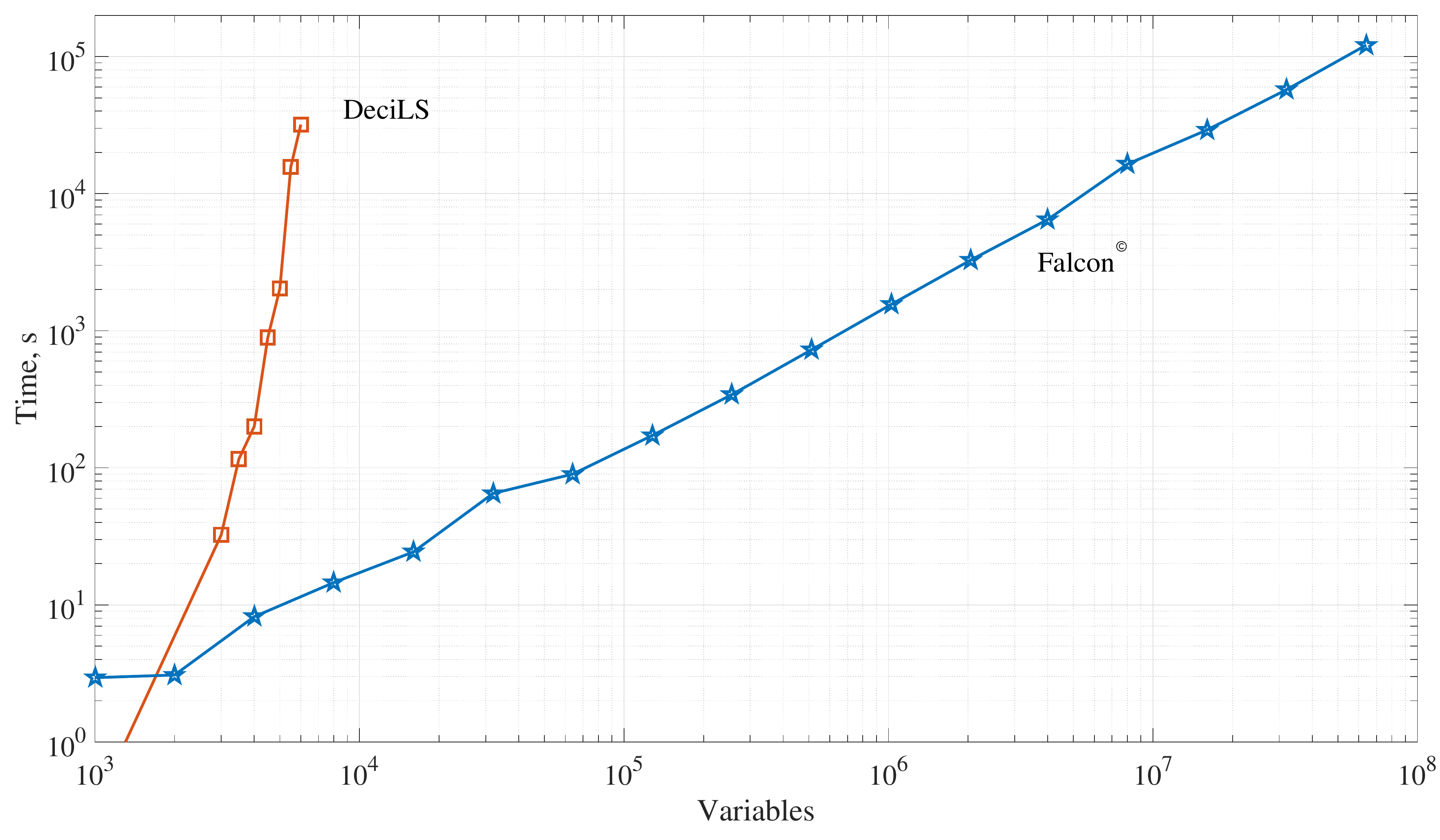}	
	\caption{Simulation time comparison between the incomplete solver DeciLS against the Falcon$^\copyright$ solver of MemComputing, Inc. for the balanced and constrained delta-Max-E3SAT. A threshold of $1.5\%$ of unsatisfiable clauses has been set. We have then tested how long DeciLS and our solver Falcon$^\copyright$ take to overcome this limit with increasing number of variables. All calculations have been performed on a single thread of an Intel Xeon E5-2680 v3 with 128 Gb DRAM shared on 24 threads. The local solver requires an exponentially increasing time to reach that limit. Our memcomputing 
		approach instead scales linearly even up to $64\times10^6$ variables (corresponding to about 1 billion literals), which required a little over than $10^5$ seconds on a single thread. We could not go beyond this limit because of memory resources (see text and Fig.~\ref{fig2}).}%
	\label{fig1}%
\end{figure*}

The DMM approach to a specific Boolean problem may be summarized as follows~\cite{DMM2,DMMperspective}:
\begin{enumerate}
\item The Boolean circuit representing the problem is constructed.
\item Traditional logic gates are replaced with \emph{self-organizing logic gates} (SOLGs)~\cite{DMM2} which are circuit elements designed to interact and collectively organize into a logically-consistent state.
\item The resulting \emph{self-organizing logic circuit} (SOLC) is fed voltages along its boundary, and allowed to evolve until it reaches equilibrium where the results of a computation may be read out.
\end{enumerate}

A detailed account of SOLGs may be found in \cite{DMM2} (see also Ref.~\cite{DMMperspective}). In essence, they may be understood as dynamical components whose equilibrium points encode the truth table of a logic gate, and can self-organize into a consistent logical 
relation of such truth table 
{\it irrespective} of the terminal to which a given truth value (a literal) is assigned. Each terminal is equipped with a \emph{dynamic error correcting module} which reads the states of its neighbors, and sets the local current and voltage to enforce logical constraints. These elements are specially designed so that the resulting dynamical system is point dissipative~\cite{hale_2010_asymptotic} (in the context of functional analysis), avoids periodic orbits~\cite{noperiod} and the chaotic behavior~\cite{no-chaos} that is typical of other non-linear dynamical systems, and utilizes components with memory to allow gates to efficiently correlate and {\it collectively} transition between states~\cite{topo} (see also below).  

The dynamics of the system are {\it deterministic} and, since they allow for collective transitions of large numbers of variables, they are also {\it non-local} in the same sense that complete combinatorial solvers are. 
Therefore, our approach contrasts sharply with those based on a stochastic local search such as simulated annealing~\cite{kirkpatricksimanneal}, and many incomplete solvers~\cite{gomes2008satisfiability,kautz2009incomplete}.  We will comment further on the reasons for these differences and the role of long-range order later.

The procedure above may be followed to construct a hardware solver for a wide variety of combinatorial and optimization problems. However, since memcomputing machines are made of non-quantum elements, the resulting behavior is also captured by a set of nonlinear ordinary differential equations describing the circuit 
(see Ref.~\cite{DMM2} for an example of DMMs equations of motion). These equations can be efficiently simulated, constituting an \emph{algorithm} for the same problem.  While this approach may seem indirect, as already noted, surprisingly the simulation of these circuits using standard numerical packages~\cite{MatLab} is sufficient to outperform the state-of-the-art combinatorial approaches on many benchmarks~\cite{exponential2017speedup,AcceleratingDL,DMMperspective}.
	
In subsequent sections we demonstrate this and stress test these simulations by showing the results of a direct comparison on the same hardware between the Falcon$^\copyright$ solver of MemComputing, Inc. and the solver DeciLS, an improved version of one of the best solvers of the 2016 Max-SAT competition~\cite{decils}. In all cases our simulations considerably outperform the other method tested, by orders of magnitude. Our solver indeed scales {\it linearly} in time and memory compared to the expected exponential scaling of the other solver.  
\begin{figure*}[ht]
	\includegraphics[width=1.8\columnwidth]{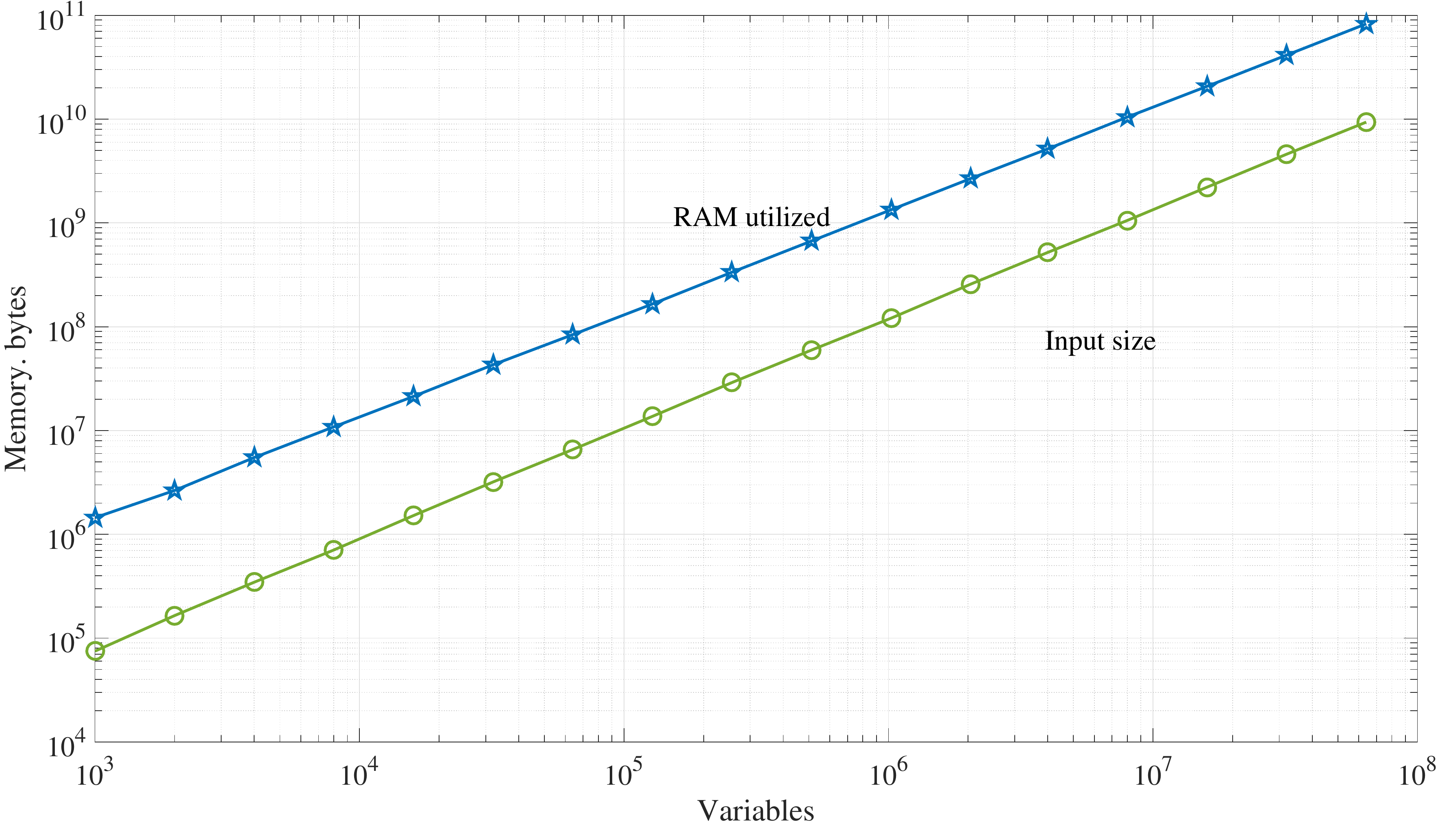}	
	\caption{Memory requirements of the Falcon$^\copyright$ solver as a function of variables for the delta-Max-E3SAT. We provide both the input size memory (open circles) and the RAM used during computation.}%
	\label{fig2}%
\end{figure*}

\section{Max-SAT}

As outlined above, we formulate our instances as Max-E3SATs, in which each clause contains exactly 3 literals and which has an inapproximability gap.  A particular instance of Max-E3SAT may be characterized by the number of variables, $N$, and the number of clauses, $M$, or alternatively the clause density, $\rho = M/N$.  As $\rho$ increases, the relationships between the variables become increasingly constrained, and it becomes less likely that an assignment satisfying all clauses will exists, i.e., the problem is more likely UNSAT.  

In the case of Random-3SAT, in which clauses are uniformly drawn from the set of possible clauses amongst $N$ variables, instances undergo a SAT/UNSAT transition at $\rho\approx 4.3$, below which an instance is almost certainly satisfiable, and above which it is almost certainly unsatisfiable (the transition being sharply defined as $N\to\infty$).  Different methods of generating instances however, will lead to different transitions: Random-3XORSAT, in which clauses of 3 literals are formed using the exclusive OR, $\oplus$, undergoes a SAT/UNSAT transition at $\rho \approx 0.918$~\cite{ricci2001simplest}.

To stress test the capabilities of simulating DMMs we utilized a set of Max-E3SAT instances based on a variant of Random-3XORSAT in which each variable was constrained to occur the same number of times (or as nearly as possible while satisfying the specified $N$ and $\rho$)~\cite{exponential2017speedup}.  This particular set of instances were chosen for their low inter-instance variability in difficulty due to the constraint placed on variable occurrences, and the relevance of MAX-XORSAT to important problems in decoding~\cite{ricci2001simplest,jia2004,cocco2006} The balanced structure of the resulting SAT makes them difficult for local solvers since, whenever a variable assignment is changed, it flips the state of all XORSAT clauses in which it was included.  Thus, in order to find transitions which satisfy a larger number of clauses, the system is forced to flip increasingly large numbers of literals concurrently.

To generate these hard instances, we first generated a random 3XORSAT instance with $\rho=1.25$ in which each variable was allowed to occur 3 or 4 times.  These were then converted to a SAT instance by replacing each XORSAT clause with 4 CNF clauses which reproduce the truth table of the XORSAT clause.  The resulting Max-E3SAT instances have a clause density of $\rho=5$. We call this problem delta-Max-E3SAT~\cite{exponential2017speedup}.

To clearly show the superior performance of our approach compared to standard algorithms, we have set a threshold of $1.5\%$ of unsatisfiable clauses and tested how long DeciLS and Falcon$^\copyright$ take to overcome this limit with increasing number of variables. DeciLS is a compiled code that combines a unit propagation based decimation (Deci) and local search (LS) with
restarts~\cite{decils}. Past useful assignments are used to break conflicts in the unit propagation and may be considered ``messages'' from the past. 

The solver Falcon$^\copyright$ is a sequential MATLAB code 
that integrates forward the DMMs equations of motion using an explicit Euler method. All calculations have been performed on a single core of an Intel Xeon E5-2680 v3 with 128 GB DRAM. As expected the local solver requires an exponentially increasing time to reach that limit already at a few thousand variables (see squares symbols in Fig.~\ref{fig1}). Instead, the simulations of DMMs scale linearly in time 
up to $64\times 10^6$ variables, which at a density of 5, corresponds to $320\times 10^6$ clauses, or about 1 billion literals. The largest case required a little 
over than $10^5$ seconds to complete. In fact the time was not a limitation, rather the memory of the processor. 

In Fig.~\ref{fig2} we present the memory used in the computation by Falcon$^\copyright$ as a function of variables. It is clear that also the memory scales linearly up to $64\times10^6$ variables. For comparison, we also display the memory size of the input file, showing that the MATLAB implementation of DMMs equations of 
motion has about an order of magnitude overhead in memory. Since the memory of the Intel Xeon used for these simulations 
was only 128 GB DRAM, that limit has set a hard stop to the actual size we could fit on that processor. Of course, different implementations (e.g., using a 
compiled language rather than an interpreted one), different hardware, etc. may permit execution of even larger instances.  

\section{Collective dynamics of DMMs}

The simulations surveyed in this paper and those already performed in relation to other problems~\cite{DMMperspective} indicate that \emph{collective} dynamics play an important role in the ability of DMMs to find good approximations to the optimum in non-convex landscapes. This contrasts sharply with the dynamics of stochastic-local-search solvers, and an analogy may be drawn to the co-tunneling events observed in quantum-annealing devices such as those manufactured by D-Wave~\cite{DenchevDWave}, although DMMs are non-quantum machines. 
 
Of particular interest is the fact that the transient dynamics of a DMM proceeds via an {\it instantonic phase} where the machine rids itself of logical defects~\cite{topo,Bearden}. Instantons are families of trajectories of the non-linear equations of motion of DMMs, and connect different critical points in the phase space that have different indexes, namely different number of unstable directions. 

In mathematical terms, if 
\begin{eqnarray}
\dot x(t) = F(x(t))\label{SDE}
\end{eqnarray}
is the equation of motion describing a DMM, with $x$ the set of elements (e.g., voltages, currents and internal state variables), and $F$ the 
flow vector field, then instantons are deterministic trajectories 
\begin{eqnarray}
\dot x_{cl}(t,\sigma) = F(x_{cl}(t,\sigma));\;\;\;\;\;\;  x_{cl}(\pm\infty,\sigma)=x_{a,b},\label{instanton}
\end{eqnarray}
that connect two arbitrary critical points of $F$, say $x_b$ and $x_a$. The parameters $\sigma$ are the so-called modulii of instantons, and 
encode their non-locality. 

Indeed, the presence of instantons creates long-range order in the system, both in time and in space~\cite{topo}. Spatial long-range order means that logic gates can 
correlate at arbitrary distances from each other. Temporal long-range order implies that the system can follow different paths to obtain the same 
solution. 

It is the presence of these instantons that renders these machines efficient in the solution search of complex problems. In fact, by transforming 
a combinatorial or optimization problem (a Boolean problem) into a physical (dynamical) one as described above, its simulation is also efficient. The reason is that the original Boolean 
problem is no longer solved combinatorially, as it is usually done. A combinatorial approach requires a search in a space that grows exponentially for a complex problem. Rather, the solution search is accomplished, via instantons, by a physical, dynamical system and the latter can be simulated 
efficiently by solving its corresponding differential equations. In addition, since instantons are topological objects, the solution search is robust against noise 
and structural disorder, a fact that was also shown explicitly in Ref.~\cite{Bearden}.

\section{Conclusions}

The performance of digital memcomputing machines on the benchmarks presented in this paper demonstrates the substantial advantages of our approach, based on the {\it simulation} of non-linear dynamical systems, compared to traditional combinatorial ones.  While we have focused on the maximum-satisfiability problem, the methods we have illustrated readily generalize to a wide variety of combinatorial optimization problems. It would then seem that physics-based 
approaches offer a lot to the world of computing, and we believe these ideas may form the basis for the next generation of computational devices.

\section*{Acknowledgments} We sincerely thank Dr. Shaowei Cai for providing us with the binary compiled code DeciLS. We also thank Haik Manukian and Robert Sinkovits for helpful discussions. M.D. and F.L.T. acknowledge partial support from the Center for Memory Recording Research at UCSD. M.D. and F.S. acknowledge partial support from MemComputing, Inc. All calculations reported here have been performed by one of us (P.C.) on a single processor of the Comet cluster of the San Diego Supercomputer Center, which is an NSF resource funded under award \#1341698. The authors would be delighted to provide, upon request, all instances of the constrained delta-Max-E3SAT used to generate Figs. 1 and 2.\\

\bibliographystyle{IEEEtran}
\bibliography{SUSYref} 

\end{document}